# CovidGAN: Data Augmentation Using Auxiliary Classifier GAN for Improved Covid-19 Detection

**ABDUL WAHEED[1], MUSKAN GOYAL[1], DEEPAK GUPTA[1], ASHISH KHANNA[1], FADI AL-TURJMAN[2], AND PLÁCIDO ROGERIO PINHEIRO[3,4], (Member, IEEE)**

[1]Maharaja Agrasen Institute of Technology, New Delhi 110086, India
[2]Artificial Intelligence Department, Research Center for AI and IoT, Near East University, 99138 Mersin, Turkey
[3]State University of Ceará, Fortaleza 60714903, Brazil
[4]University of Fortaleza, Fortaleza 60811905, Brazil

Corresponding author: Deepak Gupta (deepakgupta@mait.ac.in)

**ABSTRACT** Coronavirus (COVID-19) is a viral disease caused by severe acute respiratory syndrome coronavirus 2 (SARS-CoV-2). The spread of COVID-19 seems to have a detrimental effect on the global economy and health. A positive chest X-ray of infected patients is a crucial step in the battle against COVID-19. Early results suggest that abnormalities exist in chest X-rays of patients suggestive of COVID-19. This has led to the introduction of a variety of deep learning systems and studies have shown that the accuracy of COVID-19 patient detection through the use of chest X-rays is strongly optimistic. Deep learning networks like convolutional neural networks (CNNs) need a substantial amount of training data. Because the outbreak is recent, it is difficult to gather a significant number of radiographic images in such a short time. Therefore, in this research, we present a method to generate synthetic chest X-ray (CXR) images by developing an Auxiliary Classifier Generative Adversarial Network (ACGAN) based model called CovidGAN. In addition, we demonstrate that the synthetic images produced from CovidGAN can be utilized to enhance the performance of CNN for COVID-19 detection. Classification using CNN alone yielded 85% accuracy. By adding synthetic images produced by CovidGAN,the accuracy increased to 95%. We hope this method will speed up COVID-19 detection and lead to more robust systems of radiology.

**INDEX TERMS** Deep learning, convolutional neural networks, generative adversarial networks, synthetic data augmentation, COVID-19 detection.

## I. INTRODUCTION

Coronavirus disease is a respiratory disease caused by severe acute respiratory syndrome coronavirus 2 (SARS-CoV-2). COVID-19 was initially detected in Wuhan, China, in December 2019, and has spread worldwide since then leading to the ongoing 2020 coronavirus pandemic. More than 4.18 million cases and 286,000 deaths have been registered in more than 200 countries and territories as of 12 May 2020. Since no vaccines or cures exist, the only efficient way of human protection against COVID-19 is to reduce spread by prompt testing of the population and isolation of the infected individuals.

Certain health symptoms combined with a chest X-ray can be used to diagnose this infection. A chest X-ray can be used as a visual indicator of coronavirus infection by the radiologists. This led to the creation of numerous deep

The associate editor coordinating the review of this manuscript and approving it for publication was Shuihua Wang.

learning models, and tests have shown that it is highly likely that patients with COVID-19 infection are detected correctly by using chest radiography images.

Convolutional neural networks (CNNs) have attained state-of-the-art performance in the field of medical imaging, given enough data [1]–[4]. Such performance is accomplished by training on labeled data and fine-tuning its millions of parameters. CNNs can easily overfit on small datasets because of the large number of parameters, therefore, the efficiency of generalization is proportional to the size of the labeled data. With limited quantity and variety of samples, the biggest challenge in the medical imaging domain is small datasets [5]–[7]. The medical image collection is a very expensive and tedious process that requires the participation of radiologists and researchers [6]. Also, since the COVID-19 outbreak is recent, sufficient data of chest X-ray (CXR) images is difficult to gather. We propose to alleviate the drawbacks by using synthetic data augmentation.

Data augmentation methods are employed to extend the training dataset artificially. Current data augmentation techniques use simple modifications to incorporate affinity like image transformations and color adjustments, such as scaling, flipping, converting, improving contrast or brightness, blurring, and sharpening, white balance, etc [8]. This classical data augmentation is fast, reliable, and easy. However, in this augmentation, the changes are limited because it is structured to turn an existing sample into a slightly altered sample. In other words, classical data augmentation does not produce completely unseen data. A modern, advanced form of augmentation is synthetic data augmentation which overcomes the limitations of classical data augmentation. Generative Adversarial Network (GAN) is one such innovative model that generates synthetic images. It is a powerful method to generate unseen samples with a min-max game without supervision [9]. The general concept of the GANs is to use two opposing networks (G(z) and D(x)), where one (G(z) generator) produces a realistic image to trick the other net that is equipped to better discriminate between the true and false images (D(z) discriminator). The aim of the generator is to minimize the cost value function V(D, G) whereas the discriminator maximizes it [10]. Related works and contributions are discussed below.

### A. RELATED WORKS

Recently, the GAN framework is used by many medical imaging techniques. Zhao *et al.* [11] developed a multi-scale VGG16 network and a DCGAN based model, Forward and Backward GAN (F&BGAN) to generate synthetic images for lung-nodules classification. Beers *et al.* [12] trained a PGGAN (progressively grown generative adversarial network) to synthesize medical images of fundus pictures showing premature retinopathic vascular pathology (ROP) and glioma multimodal MRI pictures. Dai *et al.* [13] applied GAN in order to produce segmented images of lung and heart from chest X-ray. A patch-based GAN was developed in Nie *et al.* [14] to convert brain CT pictures to the corresponding MRI. They also suggested an automated image optimization model. Xue *et al.* [15] suggested two GAN networks called a Segmentor and a Critic which studied the connection between a binary brain tumor segmentation map and brain MRI pictures. Schlegl *et al.* [16] utilized GAN to study the data distribution of healthy tissue using patches in the retinal region. The GAN was then checked for anomaly detection in retinal images on patches of both unseen and healthy imagery.

The lack of data in medical imaging led us to explore ways of expanding image datasets. In the present research, we are focusing on improvements in COVID-19 detection. In order to synthesize standard CXR images, we developed an Auxiliary Classifier Generative Adversarial Network (ACGAN) based model.

In a research published in the journal Radiology [17], chest radiography outperformed laboratory testing in the detection of 2019 novel coronavirus disease. The frequency of anomalies in radiographical images rapidly increased after the onset of symptoms and peaked during the days of illness. The researchers in [18], [19] concluded that chest radiography should be used as the main COVID-19 screening method (also known as SARS-CoV-2). In particular, CXR imaging has various advantages like readily available and accessible, easily portable, and it helps in the rapid prioritization of COVID-19 suspected patients. Since the pandemic is recent, there are only a limited number of CXR images available for study. Therefore, we develop CovidGAN to generate artificial training data for CNN. The generation of artificial data is very effective in the case of small datasets and when the data includes sensitive information. GANs can synthesize images from scratch from any specific category and can produce satisfactory results when combined with other methods.

Many research projects and innovations related to COVID-19 have been proposed [20]–[22]. This paper, however, to the best of our knowledge is the first one to present a GAN architecture for improvement in COVID-19 detection.

### B. CONTRIBUTIONS

In this study, we use CNN for COVID-19 detection. CNNs are extensively used in the field of computer vision. These are widely used to examine visual imagery, and also work in image classification. In the last few years, several studies of medical imagery have applied CNNs and have recorded better performance [6]. We combine synthetic CXR images generated using CovidGAN with our proposed CNN architecture. This research has the following contributions:

1) Propose an Auxiliary Classifier Generative Adversarial Network (ACGAN) based GAN, called CovidGAN for the generation of synthetic CXR images.
2) Design a CNN-based model for COVID-19 detection.
3) Using CovidGAN for augmentation of training dataset with the CNN model for improved detection of COVID-19.

The remaining transposition is as follows. Section II defines the dataset and the CNN architecture for COVID-19 detection. Section III elaborates on the method of synthetic data augmentation for the extension of the dataset. Section IV and V show the results and conclusion of this research, respectively. Section VI discusses the limitations of this study.

## II. COVID-19 DETECTION

This section describes the characteristics of the dataset used and the CNN architecture for COVID-19 detection.

### A. DATASET GENERATION

The dataset is composed of 1124 CXR images. More precisely, there are 403 images of COVID-CXR and 721 images of Normal-CXR. To generate the dataset we collected the images from three publicly accessible datasets: 1) IEEE Covid Chest X-ray dataset [23] 2) COVID-19 Radiography Database [24] and 3) COVID-19 Chest X-ray Dataset

Initiative [25]. The decision to develop the dataset on these three datasets is driven by the fact that all of them are open-sourced and completely available to the public and research communities. The collected images are merged and the duplicate images are removed from the dataset. Image Hashing method is used to remove the duplicate images. This method creates a hash value that uniquely identifies an input image based on the contents of an image.

The most striking trend is the limited number of cases and associated CXR pictures of COVID-19 that indicate the scant availability of COVID-19 data in the public domain. Samples of the dataset are given in Fig. 4 A.

### B. CNN ARCHITECTURE

In this research, a VGG16 network is used for COVID-19 detection. A VGG16 architecture consists of twelve 3 × 3 convolutional layers. Some convolutional layers are followed by the max-pooling layer and finally, it has three fully-connected layers in the end. The stride is fixed to 1 pixel in all the convolutional layers. The five max-pooling layers use a fixed stride of 2 and a 2 × 2 pixel filter. A padding of 1 pixel is done for the 3 × 3 convolutional layers. All the layers of the network use ReLU as the activation function. The advantage of VGG16 is its simplicity and depth. Its deep architecture extracts features with low spatial resolution and give good results on image classification problems.

Our CNN uses VGG16 architecture which is connected with four custom layers at the end. A global average pooling layer is followed by a 64 units dense layer and a dropout layer with 0.5 probability. Lastly, a softmax layer is attached to find the prediction of the network.

The dataset consists of 932 training samples (COVID-CXR: 331 images and Normal-CXR: 601 images) and 192 testing samples (COVID-CXR: 72 images and Normal-CXR: 120 images). The image preprocessing steps involved are resizing and normalizing. Since the scale of the images varies in the dataset, all the images are resized to 112 × 112 × 3, using image processing SciKit. Further, each image is normalized by rescaling the pixels from [0, 255] to [0, 1]. Our CNN gets a fixed size CXR image of 112 × 112 × 3.

Since a VGG16 network has a million parameters, it requires a lot of training data and computing resources. Therefore, fine-tuning is performed to modify the parameters of the pre-trained VGG16 model so that it can adapt to the new task in hand. The custom layers of the model are trained, without updating the weights of VGG16 layers. Thus, fine-tuning updates the weights of the custom layer. This allows the new output layers to learn to interpret the learned features of the VGG16 model; which is achieved by setting the "trainable" property on each of the VGG layers to False before training.

***Training and implementation details:*** Adaptive Moment Estimation (Adam) [26] is used as the optimizer and categorical_cross_entropy as the loss function. Adam is a method for stochastic optimization which calculates adaptive learning rates for parameters. ReLU is used as the activation function of the network. The proposed CNN had approximately 14 million parameters. The learning rates of Adam are controlled by the parameter beta. The hyperparameters used for training are learning_rate = 0.001, beta = 0.9, and batch_size = 16. The network is trained for 25 epochs and after training, 85% accuracy is achieved. The proposed architecture is trained and tested using Keras deep learning library.

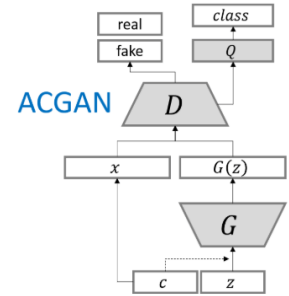

**FIGURE 1.** ACGAN Architecture.

## III. GENERATING SYNTHETIC IMAGES

The major drawback in the above CNN model was the small dataset. To extend the training data and boost the results of COVID-19 detection, we increased the data by synthetic augmentation. This section elaborates the method of augmentation in detail.

### A. AUXILIARY CLASSIFIER GENERATIVE ADVERSARIAL NETWORK (ACGAN)

Generative Adversarial Networks (GANs) utilizes two neural networks that compete with one another to create new virtual instances of data which can be transmitted as real data [10]. GANs are extensively used for image generation. In this research, we use a version of GAN called Auxiliary Classifier GAN to perform data augmentation.

GANs find it difficult to generate high-resolution samples from highly variable data sets. Conditional GAN (CGAN) is a variant of GAN which allows the model to rely on outside information to improve the sample quality. In CGAN, a latent space point and a class label are given as input to the generator and it attempts to generate an image for that class [27]. The discriminator is provided with an image as well as a class label, and it decides if the image is true or false.

AC-GAN is a type of CGAN that transforms the discriminator to predict a particular image's class label instead of receiving it as an input. It stabilizes the training process and allows the generation of high-quality images while learning a representation that is independent of the class label [28]. ACGAN architecture is shown in Fig. 1.

ACGAN applies the associated class label $c$ and noise $z$ to each produced sample [28]. The generator $G$ utilizes both to produce $X_{fake} = G(c, z)$ images. The discriminator $D$ gives

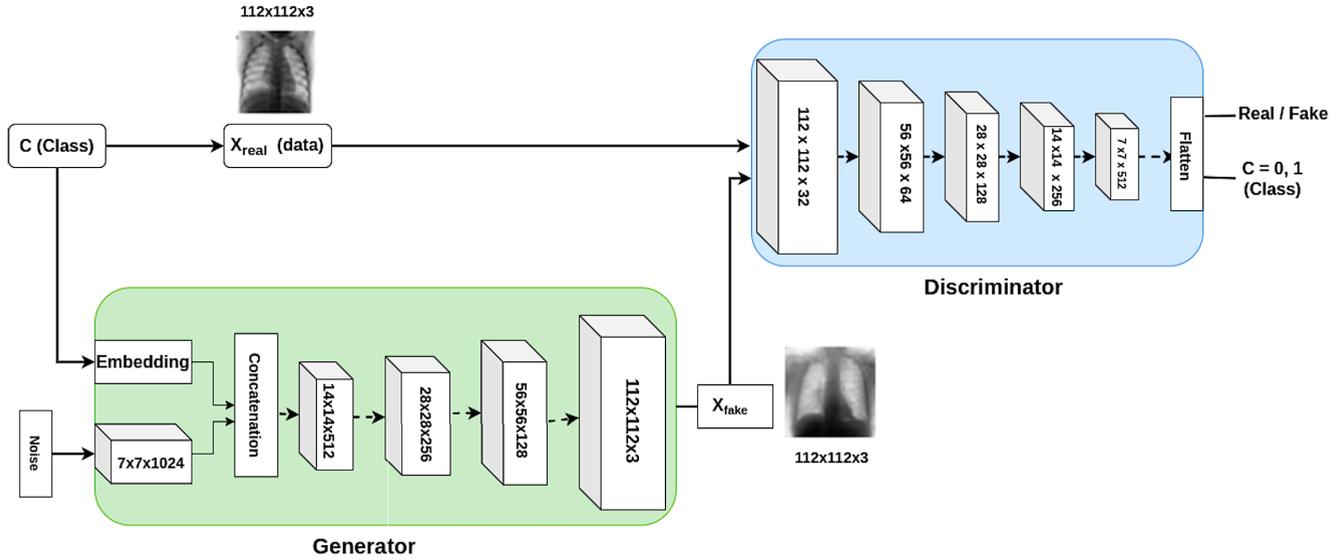

**FIGURE 2.** CovidGAN complete Architecture with generator and discriminator.

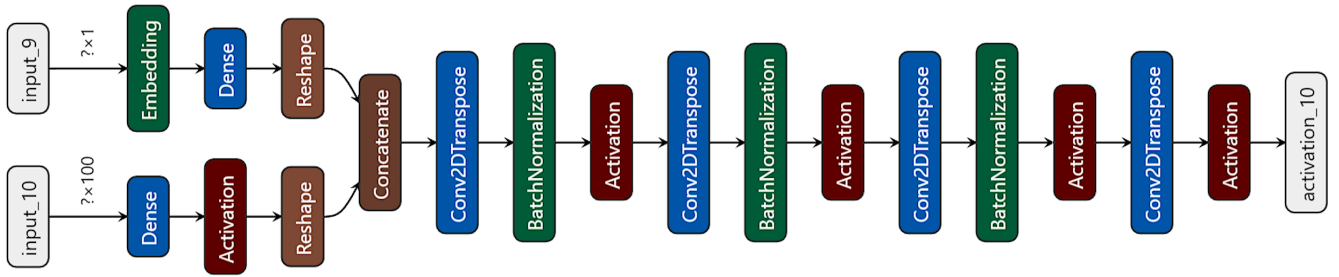

**FIGURE 3.** Layered Architecture of CovidGAN generator.

a distribution of probability over class labels and sources.

$$P(S \mid X), \; P(C \mid X) \; = \; D(X).$$

The log-likelihood of source class $L_s$ and correct class $L_c$ forms the objective function.

$$
\begin{aligned}
L_c = \; & E[\log\, P(C = c \mid X_{real})] \\
& + E[\log\, P(C = c \mid X_{fake})] \quad (1) \\
L_s = \; & E[\log\, P(S = real \mid X_{real})] \\
& + E[\log\, P(S = fake \mid X_{fake})] \quad (2)
\end{aligned}
$$

$D$ maximizes $L_s + L_c$ and $G$ maximizes $L_c - L_s$. We propose a GAN architecture based on ACGAN called CovidGAN, that produces synthetic images of CXR to improve Covid-19 detection.

### B. SYNTHETIC IMAGE AUGMENTATION USING CovidGAN

#### 1) COVIDGAN GENERATOR ARCHITECTURE

The generator takes a latent vector of noise (which is a random normal distribution with 0.02 standard deviation) and class label as input, to output a single $112 \times 112 \times 3$ image. The class label is passed through an embedding layer of 50 dimensions for categorical input. Then, it is further passed through a $7 \times 7$ node dense layer with linear activation to output a $7 \times 7 \times 1$ tensor. The point in latent space is interpreted by a $1024 \times 7 \times 7$ node dense layer to give activations that can be reshaped to $7 \times 7 \times 1024$ to get many copies of a low-resolution version of the output image. The tensors generated from class label and noise (that is $7 \times 7 \times 1$ and $7 \times 7 \times 1024$) are concatenated and passed through four transpose convolutional layers to upsample the $7 \times 7 \times 1024$ feature maps, first to $14 \times 14 \times 512$, then $28 \times 28 \times 256$, then $56 \times 56 \times 128$ and finally to $112 \times 112 \times 3$. Each transpose convolutional layer, except for the last one, is followed by a batch normalization layer and an activation layer. The model uses methodologies such as ReLU activation, a kernel of size $(5, 5)$, stride of $(2, 2)$ and a hyperbolic tangent (tanh) activation function in the output layer. The total parameters of the generator are approximately 22 million. The output of the generator is an image of shape $112 \times 112 \times 3$. Layered architecture of the generator is given in Fig. 3.

#### 2) COVIDGAN DISCRIMINATOR ARCHITECTURE

The discriminator model is a CNN architecture that has two output layers and takes one image of shape $112 \times 112 \times 3$ as input. The model outputs a prediction if the image is real

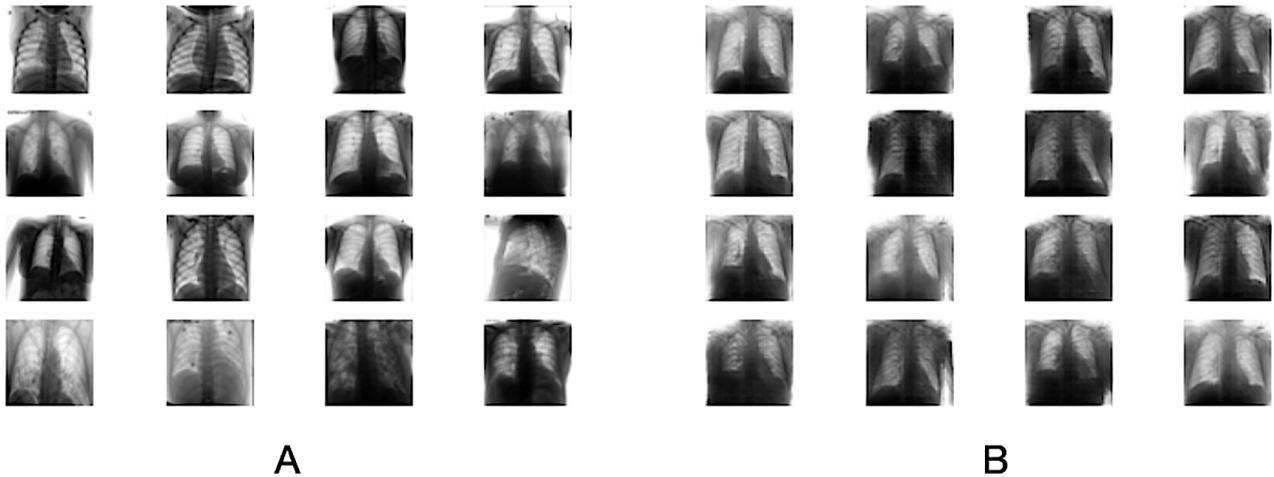

A                                                    B

**FIGURE 4.** A: Real images in dataset, B: Synthetic images generated by CovidGAN.

(class = 1) or fake (class = 0), and also outputs the class label that is COVID-CXR or Normal-CXR. Each block of discriminator represents a convolutional layer, which is followed by a batch normalization layer, an activation layer and a dropout layer with 0.5 probability. The input is downsampled from $112 \times 112 \times 32$ to $56 \times 56 \times 64$, then $28 \times 28 \times 128$, then $14 \times 14 \times 256$ and finally to $7 \times 7 \times 512$. The model uses a kernel of size (3, 3), a stride that changes alternatively from (1, 1) to (2, 2) and LeakyReLU activation function with a slope of 0.2. Discriminator has approximately 2 million parameters. The final output is flattened and the probability of the image's reality and the probability of the image belonging to each class is estimated. The first output layer with sigmoid function predicts the realness of the image. The second output layer with softmax function predicts the label.

### 3) TRAINING PROCEDURE

The generator model is stacked on top of the discriminator model. Initially, the layers of the discriminator are set as non-trainable. Thus, only the generator gets updated via the discriminator. This forms a composite model of GAN, which we call CovidGAN. The CovidGAN is trained to synthesize CXR images for both COVID-CXR and Normal-CXR class. The image preprocessing step involved resizing ($112 \times 112 \times 3$) and normalizing the images from [0, 255] to [−1, 1]. (Normalization is a process that changes the range of pixel values. Its purpose is to convert an input image into a range of pixel values that are more familiar or normal to the senses). Adam is used as the optimizer function. Adam is easy to implement, works on sparse gradients, requires little memory space, and is computationally efficient. Therefore, Adam is the best choice for the optimization of the model. The following hyperparameters are used for training CovidGAN: batch_size = 64, learning_rate = 0.0002, beta = 0.5 (beta is the momentum of Adam optimizer), number of epochs = 2000. CovidGAN has approximately 24 million parameters and it takes around 5 hours to train the model. The GAN gets optimized using two loss functions, one for each output layer of the discriminator. The first layer uses binary_crossentropy and second sparse_categorical_crossentropy. The complete architecture is trained using Keras deep learning library.

The complete CovidGAN architecture is shown in Fig. 2. The synthetic images generated from CovidGAN are shown in Fig. 4 B. CovidGAN generated 1399 synthetic images of Normal-CXR and 1669 synthetic images of COVID-CXR.

## IV. RESULTS AND DISCUSSION

In this section, we analyze the effect of synthetic data augmentation technique used for improved COVID-19 detection. Initially to perform COVID-19 detection we used the CNN classifier defined in section II. Then to improve the performance of CNN we used synthetic data augmentation technique. The performance of the model is observed on 192 testing samples (the testing samples consists of only actual data of COVID-CXR: 72 images and Normal-CXR: 120 images). We found that synthetic data augments produced (shown in Fig. 4 B) from CovidGAN enhanced the performance of CNN. An accuracy of 85% is achieved with actual data (with 0.89 precision and 0.69 recall for COVID class) that increased to 95% with synthetic augments (with 0.96 precision and 0.90 recall for COVID class). A detailed analysis is shown in Table 1. The results and Environment Setup are presented in this section.

### A. EVALUATION MEASURES AND ENVIRONMENT SETUP

The implementation of CNN and CovidGAN architecture is done using Keras [29] deep learning library. All training and testing processes are performed using Nvidia RTX 2060 GPU with 6GB memory and Intel Core i7 9th generation CPU with 16GB RAM.

We used precision, recall (or sensitivity), F1-score, and specificity to measure and analyze the performance of the CNN model using synthetic data augmentation technique. Precision is the classifier's ability to not mark a negative sample as positive and recall is the classifier's ability to

**TABLE 1.** Performance comparison for Covid-19 detection.

| Dataset | Class | Precision | Recall | F1-score | Support | Accuracy (%) | Sensitivity(%) | Specificity(%) |
|---------|-------|-----------|--------|----------|---------|--------------|----------------|----------------|
| **Actual Data (CNN-AD)** | Covid-CXR | 0.89 | 0.69 | 0.78 | 72 | **85** | **69** | **95** |
| | Normal-CXR | 0.84 | 0.95 | 0.89 | 120 | | | |
| | Macro-average | 0.87 | 0.82 | 0.84 | 192 | | | |
| | weighted-average | 0.86 | 0.85 | 0.85 | 192 | | | |
| **Actual data + Synthetic Augmentation (CNN-SA)** | Covid-CXR | 0.96 | 0.90 | 0.93 | 72 | **95** | **90** | **97** |
| | Normal-CXR | 0.94 | 0.97 | 0.96 | 120 | | | |
| | Macro-average | 0.95 | 0.94 | 0.94 | 192 | | | |
| | weighted-average | 0.95 | 0.95 | 0.95 | 192 | | | |

AD stands for actual data, SA stands for synthetic augmentation, and Support means the total number of samples.

classify all those with the disease correctly (true positive rate). F1-score is the weighted average of precision and recall. Specificity is the ability of the classifier to correctly identify those without the disease (true negative rate). In addition to total accuracy, the macro-average and weighted average are also calculated. The formulas of the measures are given below:

$$precision = \frac{TP}{TP + FP} \tag{3}$$

$$sensitivity = recall = \frac{TP}{TP + FN} \tag{4}$$

$$F1score = 2 * \frac{recall * precision}{recall + precision} \tag{5}$$

$$specificity = \frac{TN}{TN + FP} \tag{6}$$

$$Total\ accuarcy = \frac{\sum TP}{Total\ Covid19\ samples} \tag{7}$$

where $TP$ is true positives, $FP$ is false positives, and $FN$ is false negatives. Macro-average finds unweighted mean for each label without taking the label imbalance into account. Weighted average is calculated by using true instances of each label.

### B. PERFORMANCE ANALYSIS OF SYNTHETIC DATA AUGMENTATION

Table 1 analyzes the COVID-19 detection performance of CNN with synthetic data augmentation technique. We can see that when CNN is used on actual data (CNN-AD), the detection accuracy is only 85% (with 69% sensitivity and 95% specificity). As described in the previous section, we used CovidGAN data augmentation to generate synthetic images of CXR. It is observed that training CNN with actual and synthetic images (CNN-SA) yields 95% accuracy (with 90% sensitivity and 97% specificity), which is a clearly a better performance rate.

An increase in precision and recall is also recorded for both COVID (0.96 precision and 0.90 recall) and Normal (0.94 precision and 0.97 recall) class. This suggests that the synthetic augments produced have meaningful features that help in the enhancement of CNN performance.

### C. VISUALIZATION USING PCA

We use the PCA visualization and confusion matrix for analysis of the results. Principal Component Analysis (PCA) is a method which reduces the dimension of feature space such that new variables are independent [30], [31]. PCA retains large pair distances in order to optimize variance.

Steps involved in PCA:

1) **Standardization:** The mean of all the dimensions of the dataset is calculated, except the labels. The data is scaled so that each variable contributes equally to analysis. In the equation given below, z is the scaled value, x is the initial, and $\mu$ and $\sigma$ are mean and standard deviation, respectively.

$$z = \frac{x - \mu}{\sigma} \tag{8}$$

2) **Covariance Matrix Computation:** Covariance is measured between 2 dimensions. In a 3-dimensional data set (A, B, C), the covariance between the A and B dimensions, the B and C dimensions, and the A and C dimensions is measured. The covariance of two variables **X** and **Y** is computed using the following formula given below:

$$Cov(X, Y) = \frac{1}{n-1} \sum_{i=1}^{n} (X_i - X')(Y_i - X') \tag{9}$$

where $X'$ and $Y'$ are arithmetic mean of X and Y respectively, and n is number of observations. The resultant covariance matrix would be a square matrix of n x n dimensions, i.e for a 3 dimensional data the covariance matrix will be $3 \times 3$.

3) **Compute Eigenvectors and corresponding Eigenvalues:** The eigenvector and corresponding eigenvalues are computed for the covariance matrix. The corresponding eigenvalue is the factor by which the eigenvector is scaled. The eigenvector of the matrix A as a vector u that satisfies the following equation:

$$Au = \lambda u \tag{10}$$

where $\lambda$ is the eigenvalue. This means that the linear transformation is defined by $\lambda$ and the equation can be re-written as:

$$(A - \lambda I)u = 0 \tag{11}$$

where I is the identity matrix.

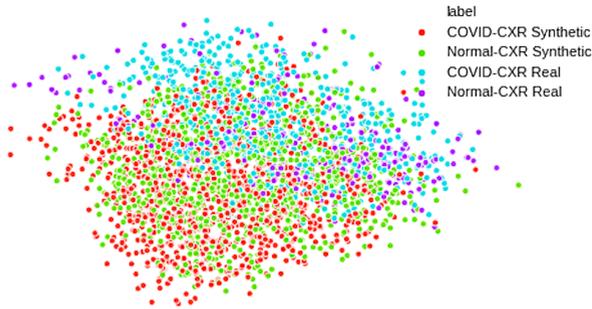

**FIGURE 5.** PCA visualization.

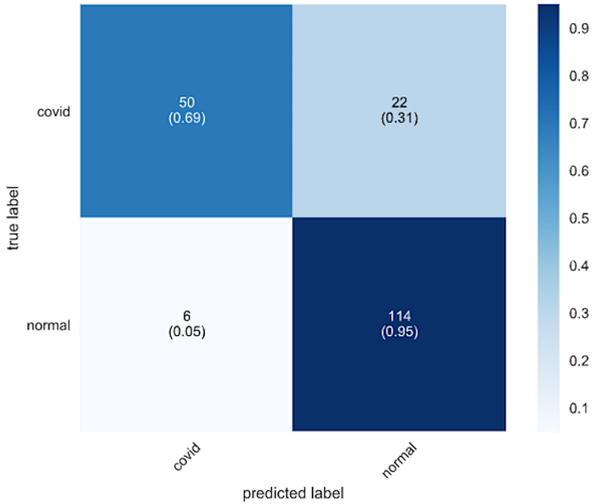

**FIGURE 6.** Confusion matrix for Covid-19 detection using CNN with actual data.

**4) Choose k eigenvectors with the largest eigenvalues:**
The eigenvectors with respect to their decreasing order of eigenvalues are sorted, $k$ is chosen out of them, where $k$ is the number of dimensions in the dataset.

**5) Recasting data along Principal Components' axes**
In the last step, our samples are transformed onto the new subspace by re-orienting data from the original axes to the ones that are now represented by the principal components.

Final Data = Feature - Vector * Transpose (Scaled (Data))

So lastly, principal components are computed and the data points in accordance with the new axes are projected.

The features or high-dimensional data are taken from the last layer of CNN. The features of the real images and synthetic images are plotted in Fig. 5. We can see that synthetic images (shown in red and green) are close to real images (shown in purple and blue).

The confusion matrices for COVID-19 detection are plotted in and Fig. 6 and Fig. 7. The confusion matrix is used to summarize the performance of the CovidGAN model. We recorded the performance of the model for 192 testing samples (COVID-CXR: 72 images and Normal-CXR: 120 images). In the dark blue colored diagonal of the matrix are the correct classifications, whereas all other entries are mis-classifications. It can be seen in Fig. 6 that 22 COVID-CXR images are misclassified as Normal-CXR (false negative) and 6 Normal-CXR images are misclassified as COVID-CXR (false positive) when CNN is trained on actual data. But in

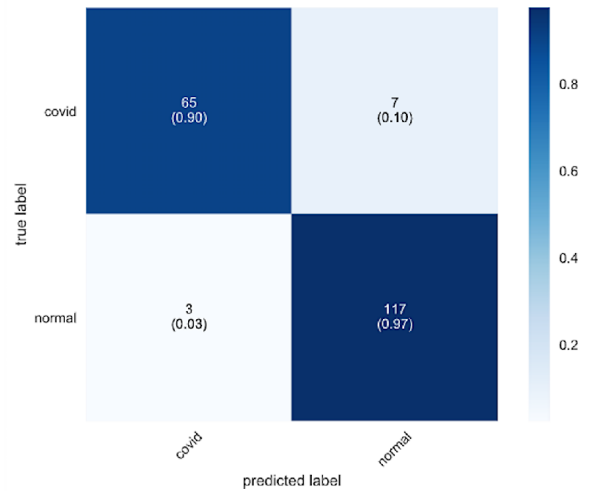

**FIGURE 7.** Confusion matrix for Covid-19 detection using CNN with synthetic data augmentation and actual data.

Fig. 7 only 7 images are misclassified as Normal-CXR when CNN is trained on actual data and synthetic augments (generated from CovidGAN). Also, the number of false positives is reduced to 3.

## V. CONCLUSION
In this research, we proposed an ACGAN based model called CovidGAN that generates synthetic CXR images to enlarge the dataset and to improve the performance of CNN in COVID-19 detection. The research is implemented on a dataset with 403 COVID-CXR images and 721 Normal-CXR images. Our limited dataset highlights the scarcity of medical images in the research communities.

Initially, the proposed CNN architecture is used to classify the two classes (that is COVID-CXR and Normal-CXR). Further, the performance of CNN with synthetic data augmentation technique is investigated.

Synthetic data augmentation adds more variability to the dataset, by enlarging it. CovidGAN is used to generate synthetic images of chest X-ray (CXR). An improvement in classification performance from 85% to 95% accuracy is recorded when CNN is trained on actual data and synthetic augments. An increase in precision and recall of both the classes are also observed.

Our findings show that synthesized images of CXR have significant visualizations and features that help in the detection of COVID-19. Lastly, a detailed analysis of the performance of our CNN architecture with synthetic data augmentation technique is given in Table 1.

In conclusion, we proposed a way to enhance the accuracy of COVID-19 detection with minimal data by generating synthetic medical images of chest X-ray. Despite its excellent results, CovidGAN is not intended to compete with laboratory testing. Instead, we hope that this approach leads to stronger and more reliable radiology systems.

In the future, we intend to improve the quality of the synthetic CXR images by training a Progressive Growing GAN [32].

## VI. LIMITATIONS

This analysis still has a variety of limitations. Firstly, GAN architecture and training can be improved further. Secondly, we used a small dataset because of the time constraints and difficulty in gathering enough data. The quality of the synthetic samples produced in this research could be improved by integrating more labeled data which improves the learning process of GAN. Thirdly, the dataset is obtained from various sources and cross-center validations were not conducted in this analysis. We have made every effort to ensure that the data collected is correctly labeled. Any mistake in data labeling, however, would probably affect the results reported. Such an impact could be especially pronounced when the dataset is small. Lastly, the only way to reliably detect COVID-19 is through medical assistance and clinical testing. The findings of this paper provide promising results that encourage the use of this approach to make more robust radiology systems. This paper also promotes a systematic large-scale gathering of COVID-CXR images.

## ACKNOWLEDGMENT

This research is dedicated to those impacted by the COVID-19 pandemic and those who are assisting in whatever way they can to fight this war. We would also like to thank doctors, nurses and all the healthcare providers who are putting their lives at risk in combating the coronavirus outbreak.

## REFERENCES

[1] S. Wang, J. Sun, I. Mehmood, C. Pan, Y. Chen, and Y. Zhang, "Cerebral micro-bleeding identification based on a nine-layer convolutional neural network with stochastic pooling," *Concurrency Comput., Pract. Exper.*, vol. 32, no. 1, p. e5130, Jan. 2020.

[2] S. Wang, C. Tang, J. Sun, and Y. Zhang, "Cerebral micro-bleeding detection based on densely connected neural network," *Frontiers Neurosci.*, vol. 13, p. 422, 2019.

[3] C. Kang, X. Yu, S.-H. Wang, D. Guttery, H. Pandey, Y. Tian, and Y. Zhang, "A heuristic neural network structure relying on fuzzy logic for images scoring," *IEEE Trans. Fuzzy Syst.*, early access, Jan. 13, 2020, doi: 10.1109/TFUZZ.2020.2966163.

[4] G. Litjens, T. Kooi, B. E. Bejnordi, A. A. A. Setio, F. Ciompi, M. Ghafoorian, J. A. W. M. van der Laak, B. van Ginneken, and C. I. Sánchez, "A survey on deep learning in medical image analysis," *Med. Image Anal.*, vol. 42, pp. 60–88, Dec. 2017.

[5] H. R. Roth, L. Lu, J. Liu, J. Yao, A. Seff, K. Cherry, L. Kim, and R. M. Summers, "Improving computer-aided detection using convolutional neural networks and random view aggregation," *IEEE Trans. Med. Imag.*, vol. 35, no. 5, pp. 1170–1181, May 2016, doi: 10.1109/tmi.2015.2482920.

[6] H. Greenspan, B. van Ginneken, and R. M. Summers, "Guest editorial deep learning in medical imaging: Overview and future promise of an exciting new technique," *IEEE Trans. Med. Imag.*, vol. 35, no. 5, pp. 1153–1159, May 2016, doi: 10.1109/tmi.2016.2553401.

[7] N. Tajbakhsh, J. Y. Shin, S. R. Gurudu, R. T. Hurst, C. B. Kendall, M. B. Gotway, and J. Liang, "Convolutional neural networks for medical image analysis: Full training or fine tuning?" *IEEE Trans. Med. Imag.*, vol. 35, no. 5, pp. 1299–1312, May 2016, doi: 10.1109/tmi.2016.2535302.

[8] A. Mikolajczyk and M. Grochowski, "Data augmentation for improving deep learning in image classification problem," in *Proc. Int. Interdiscip. PhD Workshop (IIPhDW)*, May 2018, pp. 117–122.

[9] L. Engstrom, D. Tsipras, L. Schmidt, and A. Madry, "A rotation and a translation suffice: Fooling CNN with simple transformations," 2017, *arXiv:1712.02779*. https://arxiv.org/abs/1712.02779

[10] I. Goodfellow, J. Pouget-Abadie, M. Mirza, B. Xu, D. Warde-Farley, S. Ozair, A. Courville, and Y. Bengio, "Generative adversarial nets," in *Proc. Adv. Neural Inf. Process. Syst.*, 2014, pp. 2672–2680.

[11] D. Zhao, D. Zhu, J. Lu, Y. Luo, and G. Zhang, "Synthetic medical images using F&BGAN for improved lung nodules classification by multi-scale VGG16," *Symmetry*, vol. 10, no. 10, p. 519, 2018.

[12] A. Beers, J. Brown, K. Chang, J. P. Campbell, S. Ostmo, M. F. Chiang, and J. Kalpathy-Cramer, "High-resolution medical image synthesis using progressively grown generative adversarial networks," 2018, *arXiv:1805.03144*. [Online]. Available: http://arxiv.org/abs/1805.03144

[13] W. Dai, J. Doyle, X. Liang, H. Zhang, N. Dong, Y. Li, and E. P. Xing, "SCAN: Structure correcting adversarial network for chest X-rays organ segmentation," 2017, *arXiv:1703.08770*. [Online]. Available: https://arxiv.org/abs/1703.08770

[14] D. Nie, R. Trullo, J. Lian, C. Petitjean, S. Ruan, Q. Wang, and D. Shen, "Medical image synthesis with context-aware generative adversarial networks," in *Proc. Int. Conf. Med. Image Comput. Comput.-Assist. Intervent.*, vol. 10435, 2017, pp. 417–425.

[15] Y. Xue, T. Xu, H. Zhang, L. R. Long, and X. Huang, "SegAN: Adversarial network with multi-scale l1 loss for medical image segmentation," *Neuroinformatics*, vol. 16, nos. 3–4, pp. 383–392, Oct. 2018.

[16] T. Schlegl, P. Seeböck, S. M. Waldstein, U. Schmidt-Erfurth, and G. Langs, "Unsupervised anomaly detection with generative adversarial networks to guide marker discovery," in *Proc. IPMI*, 2017, pp. 146–157.

[17] Y. Wang, C. Dong, Y. Hu, C. Li, Q. Ren, X. Zhang, H. Shi, and M. Zhou, "Temporal changes of CT findings in 90 patients with COVID-19 pneumonia: A longitudinal study," *Radiology*, Mar. 2020, Art. no. 200843. [Online]. Available: http://10.1148/radiol.2020200843

[18] M.-Y. Ng, E. Y. Lee, J. Yang, F. Yang, X. Li, H. Wang, M.-M.-S. Lui, C. S.-Y. Lo, B. Leung, P.-L. Khong, C. K.-M. Hui, K.-Y. Yuen, and M. D. Kuo, "Imaging profile of the COVID-19 infection: Radiologic findings and literature review," *Radiol., Cardiothoracic Imag.*, vol. 2, no. 1, 2020, Art. no. e200034.

[19] C. Huang *et al.*, "Clinical features of patients infected with 2019 novel coronavirus in Wuhan, China," *Lancet*, vol. 395, pp. 497–506, Feb. 2020.

[20] L. Wang and A. Wong, "COVID-Net: A tailored deep convolutional neural network design for detection of COVID-19 cases from chest radiography images," 2020, *arXiv:2003.09871*. https://arxiv.org/abs/2003.09871

[21] L. Zhong, L. Mu, J. Li, J. Wang, Z. Yin, and D. Liu, "Early prediction of the 2019 novel coronavirus outbreak in the mainland China based on simple mathematical model," *IEEE Access*, vol. 8, pp. 51761–51769, 2020.

[22] A. Imran, I. Posokhova, H. N. Qureshi, U. Masood, S. Riaz, K. Ali, C. N. John, M. Nabeel, and I. Hussain, "AI4COVID-19: AI enabled preliminary diagnosis for COVID-19 from cough samples via an app," 2020, *arXiv:2004.01275*. [Online]. Available: http://arxiv.org/abs/2004.01275

[23] *IEEE Covid Chest X-Ray Dataset*. Accessed: Mar. 7, 2020. [Online]. Available: https://github.com/ieee8023/covid-chestxray-dataset

[24] *Covid19 Radiography Database*. Accessed: Mar. 7, 2020. [Online]. Available: https://www.kaggle.com/tawsifurrahman/covid19-radiography-database

[25] *COVID-19 Chest X-Ray Dataset Initiative*. Accessed: Mar. 7, 2020. [Online]. Available: https://github.com/agchung/Figure1-COVID-chestxray-dataset

[26] D. P. Kingma and J. Ba, "Adam: A method for stochastic optimization," Dec. 22, 2014, *arXiv:1412.6980*. [Online]. Available: https://arxiv.org/abs/1412.6980

[27] M. Mirza and S. Osindero, "Conditional generative adversarial nets," 2014, *arXiv:1411.1784*. [Online]. Available: http://arxiv.org/abs/1411.1784

[28] A. Odena, C. Olah, and J. Shlens, "Conditional image synthesis with auxiliary classifier GANs," 2016, *arXiv:1610.09585*. [Online]. Available: http://arxiv.org/abs/1610.09585

[29] F. Chollet. (2015). *Keras: Deep Learning for Humans*. [Online]. Available: https://github.com/fchollet/keras

[30] K. Labib and V. R. Vemuri, "An application of principal component analysis to the detection and visualization of computer network attacks," *Ann. Telecommun.-Ann. des télécommun.*, vol. 61, nos. 1–2, pp. 218–234, Feb. 2006, doi: 10.1007/bf03231975.

[31] S. Wold, K. Esbensen, and P. Geladi, "Principal component analysis," *Chemometrics Intell. Lab. Syst.*, vol. 2, nos. 1–3, pp. 37–52, 1987.

[32] T. Karras, T. Aila, S. Laine, and J. Lehtinen, "Progressive growing of GANs for improved quality, stability, and variation," 2017, *arXiv:1710.0196*. [Online]. Available: http://arxiv.org/abs/1710.0196